\documentclass[12pt,twocolumn]{elsart3p}
\usepackage[dvips]{graphicx,color}
\usepackage{bm}

\newcommand{\rbk}[1]{\left( #1 \right)}

\newcommand{\retn}{\nonumber \\ }

\newcommand{\dif}[2]{\frac{d #1}{d #2}}

\newcommand{\MARU}[1]{{\ooalign{\hfil#1\/\hfil\crcr\raise .167ex\hbox{\mathhexbox20D}}}}

\begin{document}

\begin{frontmatter}

\title{Molecular Dynamics Simulation of the Hydrogen Isotope Sputtering of Graphite}

\author[NAGOYA]{Atsushi ITO\corauthref{cor}}, 
 \ead{ito.atsushi@nifs.ac.jp}
 \corauth[cor]{Corresponding author. Present address: National Institute for Fusion Science, Oroshi--cho 322-6, Toki 509--5292, Japan. Tel: +81 572 58 2351; Fax: +81 572 58 2626}
\author[NIFS]{Hiroaki NAKAMURA}, 
 \address[NAGOYA]{Department of Physics, Graduate School of Science, Nagoya University, Furo--cho, Chikusa--ku, Nagoya 464--8602, Japan.}
 \address[NIFS]{National Institute for Fusion Science, Oroshi--cho 322-6, Toki 509--5292, Japan.}

\date{\today}

\begin{abstract}
We used a molecular dynamics simulation with the modified Brenner reactive empirical bond order potential to investigate the erosion of a graphite surface due to the incidence of hydrogen, deuterium, and tritium atoms.
Incident particles cause pressure on the graphite surface, and the chemical bond between graphene layers then generates heat to erode the graphite surface.
We evaluated the speed of surface destruction by calculating the pseudo--radial distribution function.
The speed of surface destruction due to incident hydrogen isotopes was higher than that due to hydrogen atoms.
The surface destruction increased exponentially and its decay time constant was a power function of the incident energy.
We measured the erosion yield, which indicated a steady state for the graphite erosion.
The erosion yield flux in the steady state increased linearly with the incident energy.
The erosion yield flux was almost independent of the type of incident particle, and the erosion yield start time was smaller for hydrogen isotopes than for hydrogen atoms.

\end{abstract}


\end{frontmatter}

\section{Introduction}

\begin{figure*}[!ht]
	\resizebox{\linewidth}{!}{\includegraphics{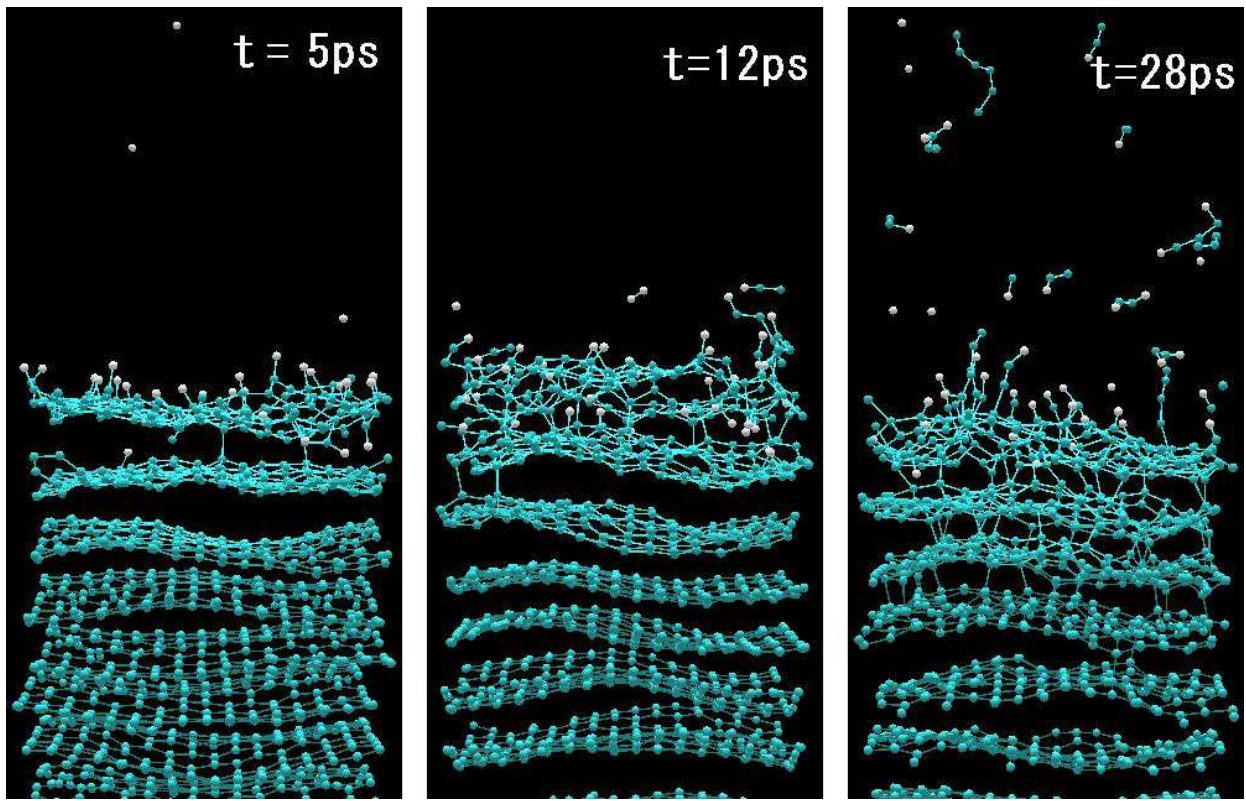}}
	\caption{The erosion of the graphite surface for deuterium incident at 5 eV.}
	\label{fig:snapshot5eV}
\end{figure*}

In the context of research into nuclear fusion, we studied the plasma surface interaction (PSI) problem \cite{Nakano,Roth,Roth2,Mech,LHD}.
A portion of the plasma confined in an experimental device falls onto a diverter wall, which is a shield made of graphite or carbon fiber composite tiles.
The incident hydrogen plasma erodes these carbon tiles in a process called chemical sputtering.
The erosion produces hydrocarbon molecules, such as $\textrm{CH}_{\mathrm{x}}$ and $\textrm{C}_2\textrm{H}_{\mathrm{x}}$, which affect the plasma confinement.

To solve the PSI problem, the mechanism of the graphite erosion has been researched using molecular dynamics simulation (MD) \cite{Salonen,Salonen2,Alman,Marian}.
Previously, we investigated the PSI of graphite surfaces using the modified Brenner reactive empirical bond order (REBO) potential \cite{Nakamura}.
That MD simulation showed that if incident energy was 5 eV, almost of all incident hydrogen atoms were absorbed by the graphite surface, while if the incident energy was 15 eV, most incident hydrogen atoms were reflected.
This absorption and reflection can be explained by the chemical reaction between a single hydrogen atom and a single graphene \cite{Ito_ICNSP, Ito_gh1}.
However, the number of absorbed hydrogen atoms seems to be independent of this graphite erosion because although the hydrogen atoms are absorbed by the first graphene on the surface only, multiple graphenes are destroyed simultaneously.
Almost all of the absorbed hydrogen atoms are located on one side of the first graphene.
The chemical bond between the first and second graphenes triggers the graphite erosion.
We assume that the momentum of the incident hydrogen atoms, which is considered to be pressure from the point of view of macroscopic thermodynamics, presses the first graphene against the second.

Nuclear fusion research must not only consider the hydrogen atom, but also the hydrogen isotopes deuterium and tritium.
It is important to understand the difference in the PSI due to these isotopes.
A study of the chemical reaction between a single graphene and a single hydrogen isotope atom showed that the absorption and reflection rates as functions of the incident energy differed from that of the hydrogen atom \cite{Nakamura_PET}.
These differences will be also be reflected in the graphite erosion.
Moreover, the incident momentum of the hydrogen isotopes differs than that of the hydrogen atoms even if the incident energies are the same.
Therefore, the pressure of hydrogen isotopes on the graphite surface is greater than that of hydrogen atoms.
If the chemical bonding between the first and second graphenes, caused by pressure from the incident momentum, triggers the graphite erosion, we would expect that the hydrogen isotopes would accelerate this erosion more than hydrogen atoms would.

We used MD simulation to investigate graphite erosion due to the incidence of hydrogen, deuterium, and tritium atoms.
We describe the simulation model and method in \S \ref{ss:SimMethod}.
In \S \ref{ss:Result}, we present and discuss the simulation results.
This paper concludes with a \S \ref{ss:Summary}.

\begin{figure*}[!ht]
	\resizebox{\linewidth}{!}{\includegraphics{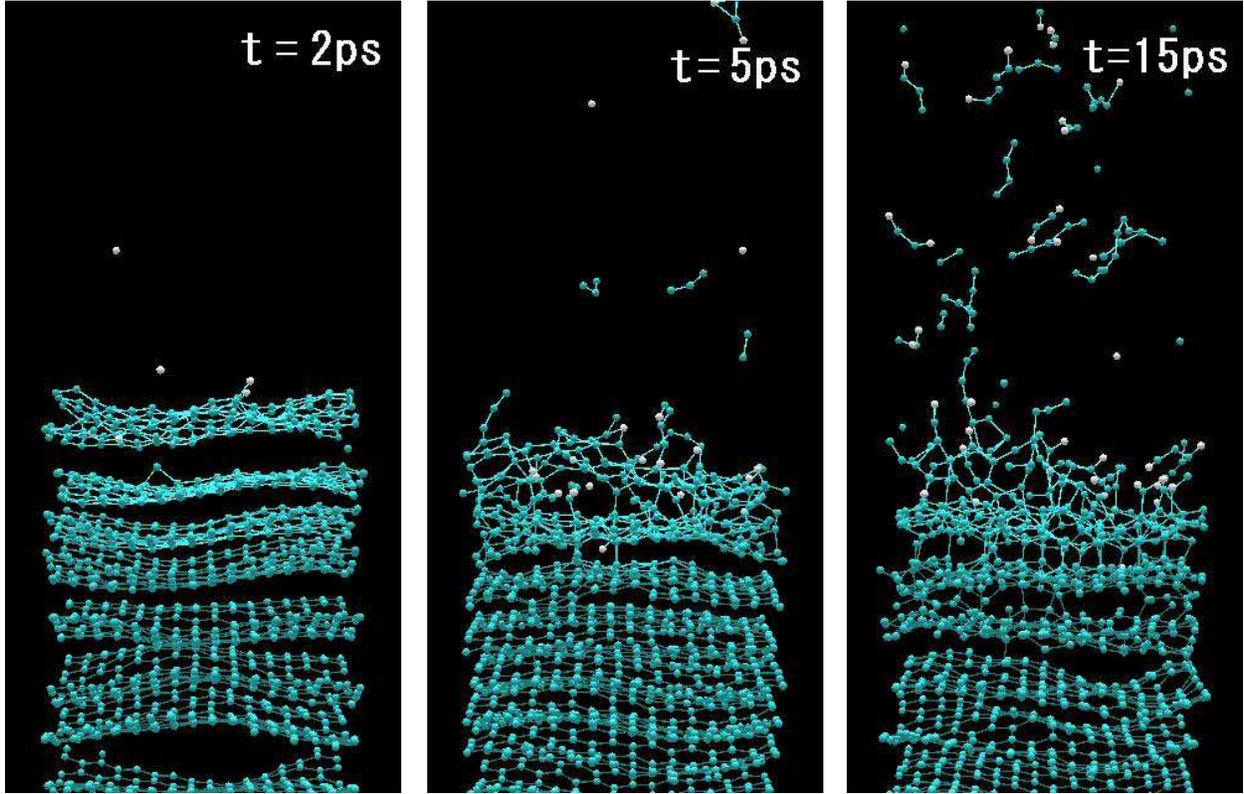}}
	\caption{The erosion of the graphite surface for deuterium incident at 15 eV.}
	\label{fig:snapshot15eV}
\end{figure*}

\section{Simulation Method}\label{ss:SimMethod}

As the graphite, we located eight graphenes \cite{Boehm} with an ``ABAB'' lattice structure parallel to $x$--$y$ plane.
Each graphene consisted of 160 carbon atoms measuring 2.13 nm $\times$ 1.97 nm.
The size of the simulation box in the $x$-- and $y$--directions equaled that of graphene with the periodic boundary condition.
The inter--layer distance of the graphite was initially 3.35 \AA.
The initial distribution of the momentum of the carbon atoms followed the Maxwell--Boltzmann distribution at 300 K.
During the simulation, only two carbon atoms were fixed to support the base of the graphite.
One was the center atom of the 7--th graphene from the surface, and the other was located at the boundary of the 8--th graphene.
The graphite surface was oriented to face the positive $z$--direction.

For the simulation, 500 or more hydrogen or isotope atoms were injected at regular time intervals of 0.1 ps.
The $x$-- and $y$--coordinates of the injection point were set at random.
The $z$--coordinate of the injection point was 60 \AA.
The initial momentum vector (0, 0, $p_0$) was parallel to the $z$--axis, and was defined by
\begin{eqnarray}
	p_0 = \sqrt{2 m E_{\mathrm{I}}}, \label{eq:inip}
\end{eqnarray}
where $E_\mathrm{I}$ is the incident energy, and $m$ is the mass of the incident particle, which is 1, 2, or 3 u for the case of hydrogen, deuterium, or tritium, respectively.

We performed our MD simulation under \textit{NVE} conditions, where the number of atoms, volume, and total energy are conserved, except for the addition of incident atoms and removal of outgoing atoms.
The simulation time was developed using second order symplectic integration \cite{Suzuki}.
The chemical interaction was represented by the modified Brenner REBO potential \cite{Ito_gh1, Brenner}:
\begin{eqnarray}
	U \equiv \sum_{i,j>i} \Bigg[V_{[ij]}^\mathrm{R}( r_{ij} )
		 - \bar{b}_{ij}(\{r\},\{\theta^\mathrm{B}\},\{\theta^\mathrm{DH}\}) V_{[ij]}^\mathrm{A}(r_{ij}) \Bigg],\retn
	\label{eq:model_rebo}
\end{eqnarray}
where $r_{ij}$ is the distance between the $i$--th and $j$--th atoms.
The functions $V_{[ij]}^{\mathrm{R}}$ and $V_{[ij]}^{\mathrm{A}}$ represent repulsion and attraction, respectively.
The function $\bar{b}_{ij}$ generates multi--body force.
To conserve the accuracy of the calculation, the time step was $5\times10^{-18} \mathrm{~s}$.

\section{Results and Discussion}\label{ss:Result}

We performed simulations for incident energies in the range 0.5--30 eV.
Figure \ref{fig:snapshot5eV} shows a snapshot of the graphite erosion due to deuterium incidence at 5 eV.
This figure clearly illustrates the erosion process.
In the initial short time period (5 ps), the deuterium atoms are generally absorbed. This is consistent with previous research findings concerning the chemical reaction between a single deuterium atom and a single graphene.
Although the first graphene absorbs many deuterium atoms, no erosion yet occurs.
When the first and second graphenes are linked by a covalent bond (12 ps), the erosion process starts, using the binding energy of the chemical bond between the graphenes.
The graphite erosion then advances to the lower graphenes connected by covalent bonds (28 ps).
For deuterium incident at 15 eV (see Fig. \ref{fig:snapshot15eV}), although the first graphene reflects almost all of the deuterium atoms (2 ps), and the chemical bonding between the graphenes results in graphite erosion (5 ps), as in the case of 5 eV.
This behavior is the same for both hydrogen atoms and hydrogen isotopes.
Therefore, we conclude that the chemical bonding between the graphenes is the dominant mechanism in the erosion process, rather than the absorption and reflection at the graphene surface.

\begin{figure}[ht]
	\begin{tabular}{c}
	\resizebox{\linewidth}{!}{\includegraphics{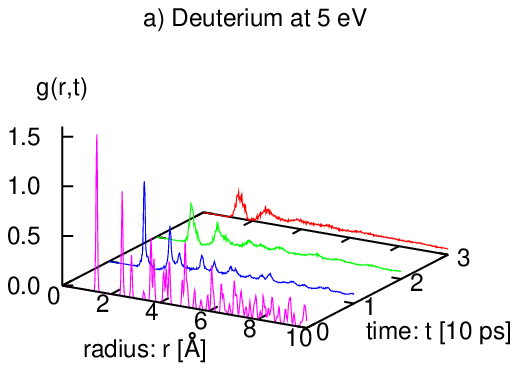}} \\
	\resizebox{\linewidth}{!}{\includegraphics{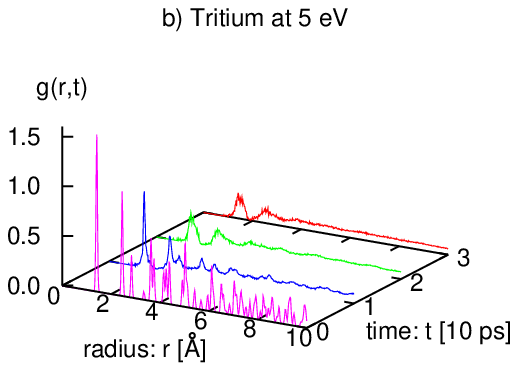}}
	\end{tabular}
	\caption{The pseudo--radial distribution function $g(r,t)$ as a function of the distance between carbon atoms $r$ and time $t$.}
	\label{fig:rdf}
\end{figure}

We investigated the surface destruction (amorphization of the graphite surface) and the erosion yield in our study of graphite erosion.

\begin{figure*}[!hp]
	\begin{tabular}{cc}
		\resizebox{0.5\linewidth}{!}{\includegraphics{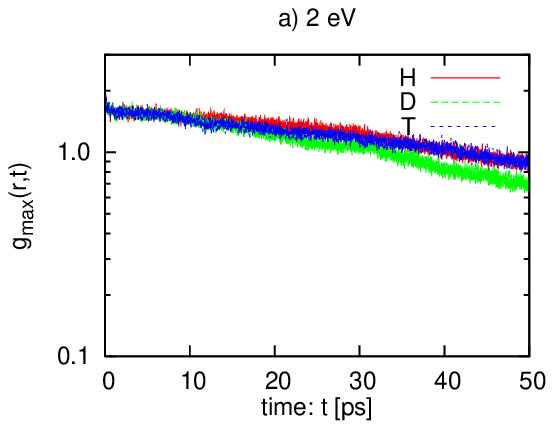}} &
		\resizebox{0.5\linewidth}{!}{\includegraphics{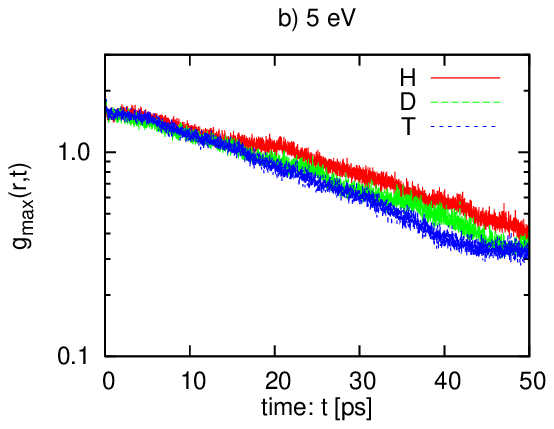}} \\
		\resizebox{0.5\linewidth}{!}{\includegraphics{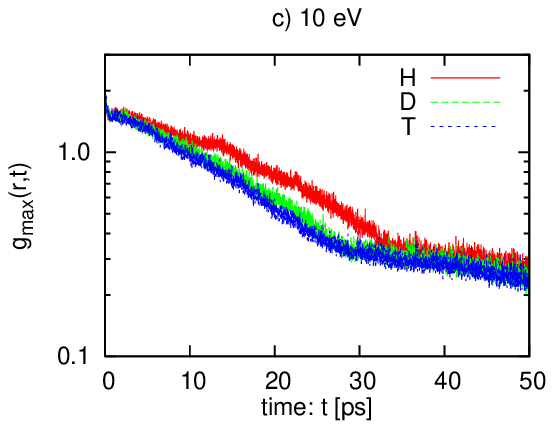}} &
		\resizebox{0.5\linewidth}{!}{\includegraphics{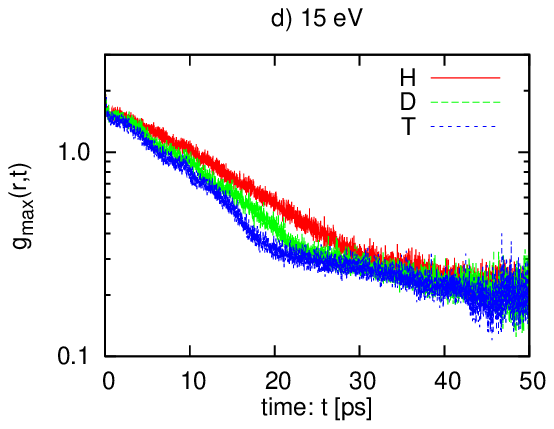}} \\
		\resizebox{0.5\linewidth}{!}{\includegraphics{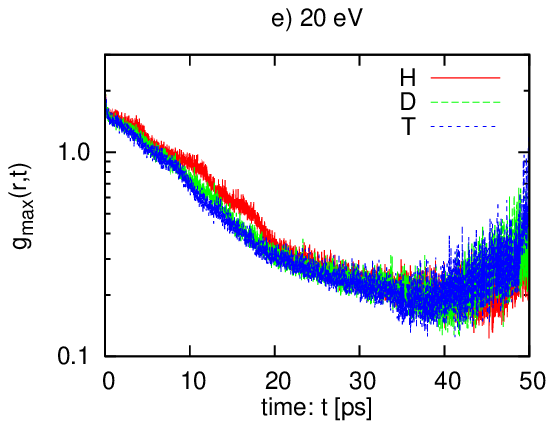}} &
		\resizebox{0.5\linewidth}{!}{\includegraphics{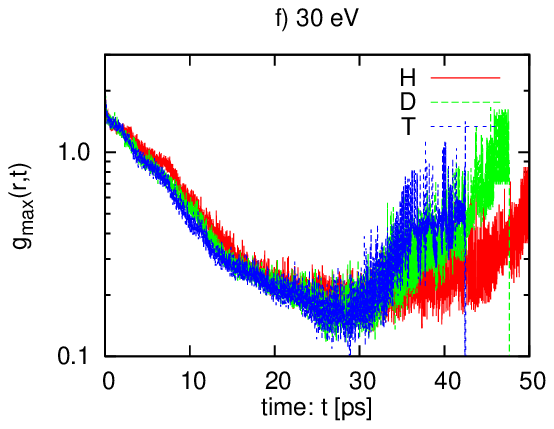}} \\
	\end{tabular}
	\caption{The maximum value of the pseudo--radial distribution function $g_max(r,t)$ as a function of the time $t$.}
	\label{fig:grmaxlog}
\end{figure*}

\begin{figure}[!ht]
	\centering
	\resizebox{\linewidth}{!}{\includegraphics{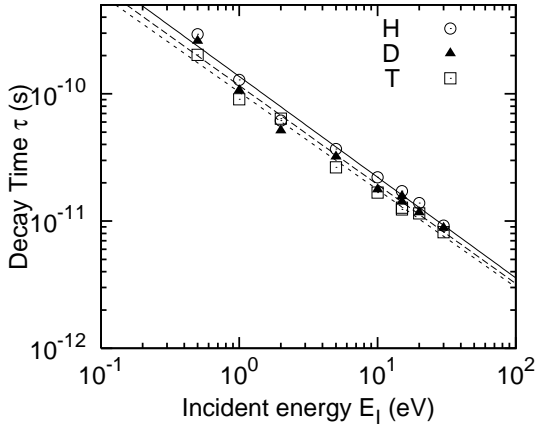}}
	\caption{The decay time constant $\tau$ as a function of the incident energy. The solid, long dashed, and short dashed lines indicate values determined by the simulation for incident hydrogen, deuterium, and tritium, respectively.}
	\label{fig:dtcEi}
\end{figure}

To estimate the surface destruction, we must first calculate the radial distribution function.
However, we cannot define the volume in this simulation model because there is no boundary in the $z$--direction.
Therefore, we use a pseudo--radial distribution function $g(r,t)$ defined by
\begin{eqnarray}
	g(r,t) \equiv \frac{1}{4 \pi r^2} \dif{n(r,t)}{r},
\end{eqnarray}
where $r$ is the distance between two particles, $t$ is time, and $n(r,t)$ is the number of carbon atoms located at a distance of less than $r$ at time $t$.
Although the pseudo--radial distribution function $g(r,t)$ differs from usual radial distribution function in that it is not normalized by the number density, it is sufficient to estimate the lattice structure because the number density is constant.
Figure \ref{fig:rdf} shows the change of the pseudo--radial distribution function $g(r,t)$ with time.
It is the same for hydrogen, deuterium, and tritium. Initially, the peaks of the graphite lattice structure are sharp, but then they broaden as time goes on, indicating the progress of the surface destruction.
To measure the speed of surface destruction, we plotted the maximum values of the pseudo--radial distribution function $g_\mathrm{max}(r,t)$ as a function of time (see Fig. \ref{fig:grmaxlog}).
These values often indicate C--C bonds of length $r = 1.42$~\AA and correspond to the number of $\textrm{sp}^2$ bonds.
From this figure, if we neglect the irregular regions, which are for $t > 35$~ps at 20 eV and for $t > 25$~ps at 30 eV, it is obvious that the number of $\textrm{sp}^2$~bonds decreases exponentially with time.
The speed of the surface destruction due to incident hydrogen isotopes is greater than that due to hydrogen atom.
Figure \ref{fig:dtcEi} shows that the decay time constant $\tau$ of the surface destruction is a power function of the incident energy $E_\mathrm{I}$.
We found that the maximum value of the pseudo--radial distribution function $g_\mathrm{max}(r,t)$ and the decay time constant $\tau$ can be represented by
\begin{eqnarray}
	g_\mathrm{max}(r,t) \propto \exp\rbk{-\frac{t}{\tau}},\\
	\tau = C {E_\mathrm{I}}^a,
\end{eqnarray}
where the scaling exponent $a$ for the hydrogen, deuterium, and tritium atoms is $-0.792$, $-0.778$, and $-0.767$, respectively. The constant value $C$ for the hydrogen, deuterium, and tritium atoms is $1.37\times10^{-10}$, $1.12\times10^{-10}$, and $1.03 \times 10^{-10}$, respectively.
This indicates that the scaling exponent $a$ is almost independent of the type of incident particle, while the constant value $C$ does depend on the incident particle type.

\begin{figure*}[!hp]
	\begin{tabular}{cc}
		\resizebox{0.5\linewidth}{!}{\includegraphics{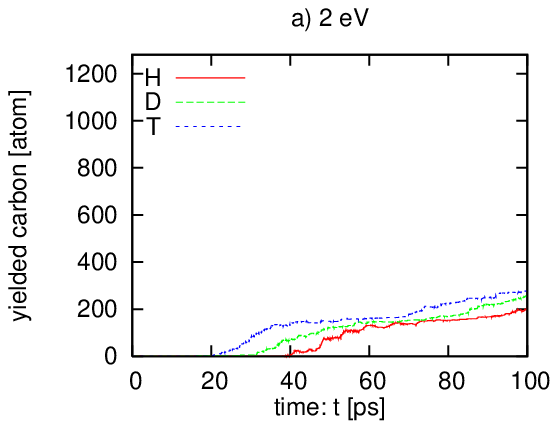}} &
		\resizebox{0.5\linewidth}{!}{\includegraphics{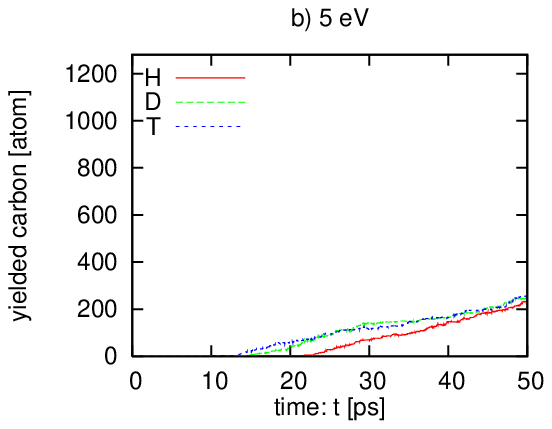}} \\
		\resizebox{0.5\linewidth}{!}{\includegraphics{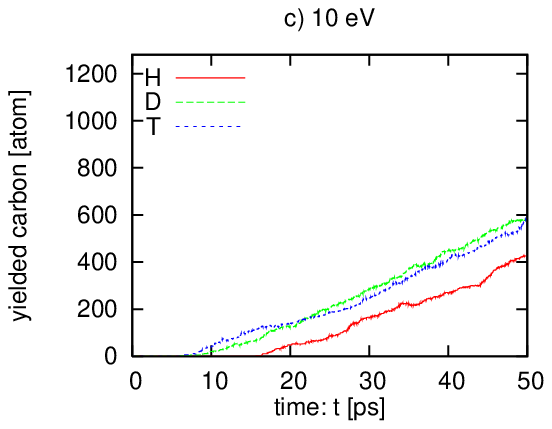}} &
		\resizebox{0.5\linewidth}{!}{\includegraphics{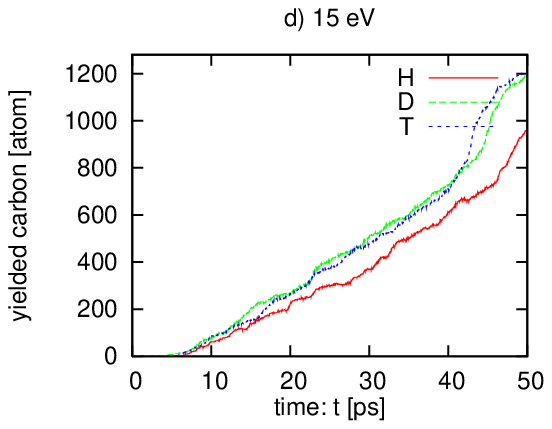}} \\
		\resizebox{0.5\linewidth}{!}{\includegraphics{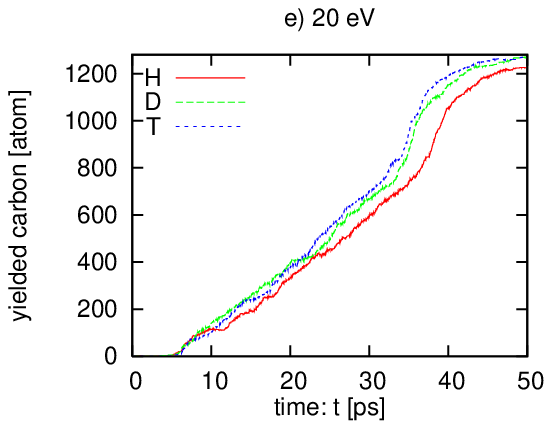}} &
		\resizebox{0.5\linewidth}{!}{\includegraphics{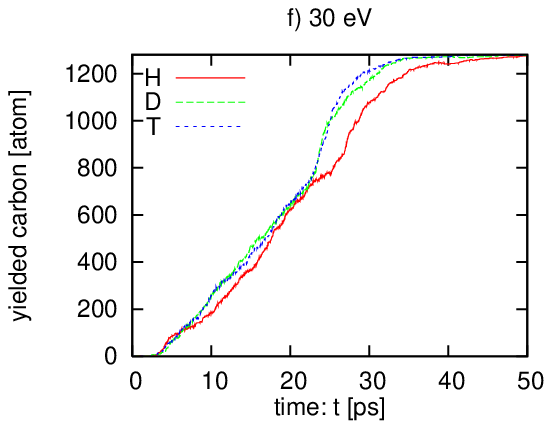}} \\
	\end{tabular}
	\caption{The erosion yields $Y$ of carbon atoms as a function of time.}
	\label{fig:yield}
\end{figure*}

\begin{figure}[ht]
	\centering
	\resizebox{\linewidth}{!}{\includegraphics{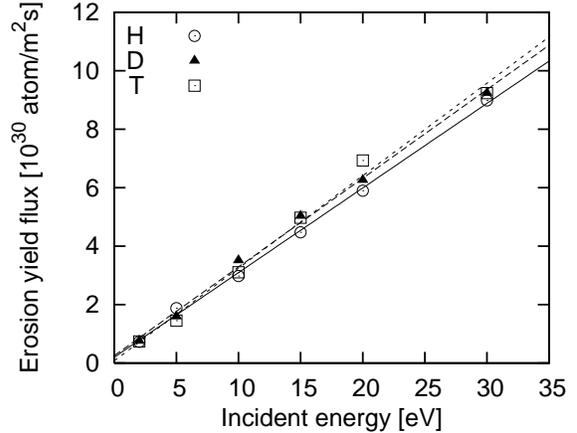}}
	\caption{The erosion yield flux $\phi$ of carbon atoms as a function of the incident energy. The solid, long dashed, and short dashed lines indicate the flux due to hydrogen, incident deuterium, and tritium, respectively.}
	\label{fig:flux}
\end{figure}

\begin{figure}[ht]
	\centering
	\resizebox{\linewidth}{!}{\includegraphics{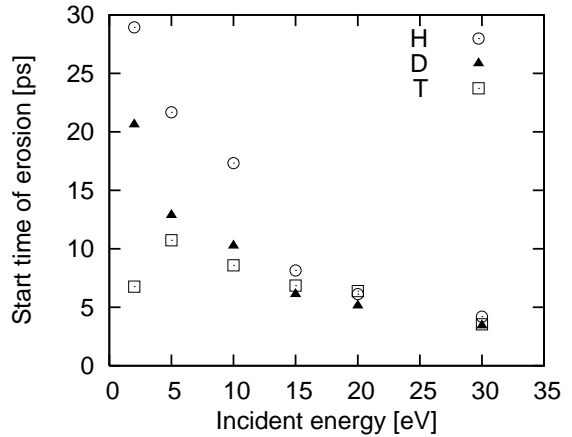}}
	\caption{The erosion yield start time $t_0$ as a function of the incident energy.}
	\label{fig:startTm}
\end{figure}

Next, we observe the erosion yield.
In our simulation, almost all of the molecules produced had a chain structure.
That is to say, carbon atoms were bounded by $\textrm{sp}^1$ bonds, and hydrogen atoms and isotopes often terminated the end of a carbon chain (e.g., H--C--C--C--C--H).
Since a chemical reaction occurs between these molecules and changes their molecular structure, we did not specify the types of molecule produced.
For the erosion yields $Y$, we counted the number of carbon atoms that moved to the region $z > 24$~\AA, where the first graphene is initially located on $z = 11.7$~\AA.
Fig. \ref{fig:yield} shows the erosion yield $Y$ as a function of time t.
First, the yield molecules are created after an initial delay.
This implies that chemical sputtering requires an increase in surface temperature.
Second, the erosion yield $Y$ increases linearly to a level of 740 carbon atoms.
This linear increase of $Y$ is thought to be the graphite erosion steady state.
After reaching 740 carbon atoms, the erosion yield $Y$ increases rapidly to the point of saturation.
It is clear from our MD simulations that the base graphene is destroyed when $Y > 740$ carbon atoms.
The eroded graphene appears to transfer heat to lower graphenes during the steady state, and the rapid increase and saturation $(Y > 740)$ are due to the absence of additional lower graphenes to absorb the heat.
We fit the erosion yields $Y$ to the following linear function $y_\mathrm{l}(t)$:
\begin{eqnarray}
	y_\mathrm{l}(t) = \phi S (t - t_0),
\end{eqnarray}
where the fitted region is $5 < Y < 740$.
The coefficient S is the square measure of 2.13 nm $\times$ 1.97 nm.
The fitting parameters $\phi$ and $t_0$ correspond to the erosion yield flux and the erosion yield start time, respectively.
Fig. \ref{fig:flux} shows the incident energy and isotope dependence of the erosion yield flux.
The erosion yield flux increases as a linear function of the incident energy, but it is almost independent of the type of isotope.
It is clear that in the steady state, the erosion yield depends on the incident energy rather than the incident momentum, which depends on the type of incident particle.
Moreover, the linear increase of the erosion yield flux indicates that the total yield on a real graphite surface also increases linearly with the incident energy.
In other experiments \cite{Mech2, Garcia}, the total yield increased linearly with the incident energy up to 50 eV.
Fig. \ref{fig:startTm} shows the incident energy and isotope dependence of the erosion yield start time.
This indicates that incident hydrogen isotopes start the erosion more quickly than hydrogen atom does.
In addition, the erosion yield start time decreases as the incident energy increases, except with tritium at 2 eV.
Our explanation of this is as follows. The erosion yield requires an increase in the graphite surface temperature, and it is the chemical bonding between the first and second graphenes that generates the heat to cause this surface temperature increase.
Because the graphenes are linked due to a pressure on surface, which is derived from incident momentum, the erosion yield start time depends on the type of incident particle.

Next, we address two other important problems.

The first is the problem of graphite interlayer interaction.
In this work, although the interaction of a covalent bond is represented by the modified Brenner REBO potential model, the graphite interlayer intermolecular force is not included.
While the interaction energy between graphite layers is thought to be about 0.01 times that of a covalent bond, its true value is not yet known \cite{Hasegawa}.
In addition, when using simple two--body interaction functions, such as the van der Waals interaction in the graphite interlayer interaction model, the real graphite ``ABAB'' structure cannot be represented.
Therefore, we do not have an intermolecular interaction model that is suitable for graphite.
However, the modified Brenner REBO potential generates a strong repulsive force when the interlayer distance is less than 2 \AA.
Although the interlayer distance cannot be kept at 3.35 \AA, the repulsive force maintains the layer structure of the graphite as long as the graphite is not subjected to a strong external force or high pressure, and its temperature is maintained below the melting point.
In his paper on the chemical sputtering of polymer, Yamashiro described the importance of the intermolecular attractive force in creating a cluster structure and bringing about thermal deposition \cite{Yamashiro}.
However, the intermolecular repulsive force, which maintains the interlayer distance and ``ABAB'' structure, plays a more important role than the intermolecular attractive force in the graphite structure.
In addition, the hydrocarbon molecules that are produced tend to prefer chain structures to cluster structures.
Therefore, we determined that it was necessary to investigate the importance of the interlayer intermolecular force in the specific case of graphite erosion.
We have been creating a new intermolecular potential model for graphite interlayer interaction.
This work is essential for comparison with our future work, which will include the graphite interlayer intermolecular interaction.

The second problem relates to temperature control.
In this work, we can achieve steady state graphite erosion without the need to cool the graphite, since heat is conducted to the lower graphenes.
In other words, the lower graphenes act as a heat reservoir.
MD simulations generally use a thermostat to provide rapid cooling because of the picosecond--scale simulation time.
Consequently, the dynamics of atoms in the simulation differ from those in reality; for example, they halt the reaction due to a trap on the local potential minimum.
Moreover, because the rate of heat conduction depends on the interaction potential model among atoms, which represents a realistic interaction, it is fairly realistic but slow, judging from the picosecond simulation time scale, and cannot be changed too quickly.
Therefore, even if the thermostat were not set to the position where a chemical reaction occurs simply to preserve the realistic dynamics of atoms, sufficient cooling would still not be possible because of the slow rate of heat conduction.
Therefore, we believe that the steady state of the graphite erosion that we have identified here is a realistic phenomenon, even if the steady state is maintained for a short time only, and the incident flux is very high.
A longer--term steady state would likely be possible using graphite with more graphene layers.

\section{Summary}\label{ss:Summary}
We have studied the graphite erosion process due to the incidence of hydrogen, deuterium, and tritium atoms using MD simulation with the modified Brenner REBO potential for incident energy $E_\mathrm{I}$ in the range 0.5--30 eV.
For $E_\mathrm{I} = 5\textrm{~eV}$, both hydrogen atoms and hydrogen isotopes were generally absorbed by the graphene surface.
At $E_\mathrm{I} = 15\textrm{~eV}$, however,  almost all of the incident particles were reflected.
When the incident particles press the first graphene against the second, erosion of the graphite surface is caused by the binding energy of the chemical bond between the graphene layers.
We evaluated the speed of the surface destruction by calculating the pseudo--radial distribution function $g(r,t)$, the peak values of which decrease with time.
The maximum value of the pseudo--radial distribution function $g_\mathrm{max}(r,t)$, which generally indicates the number of $\textrm{sp}^2$ C--C bonds, decreased exponentially.
The surface destruction due to the hydrogen isotopes progressed faster than the destruction caused by hydrogen atoms.
Moreover, we found a power law between the decay time constant and incident energy, and the scaling exponent seems to be independent of the type of incident particle.
Measuring the erosion yields indicated a steady state of the graphite erosion.
The erosion yield flux in the steady state increased linearly with the incident energy, which agreed with experimental results.
The erosion yield flux is almost independent of the type of incident isotope.
Observation of the erosion yield start time showed that the erosion due to hydrogen isotopes started earlier than the erosion due to the hydrogen atoms.

\section*{Acknowledgments}

The authors acknowledge helpful comments by and stimulating discussion with Dr. Arimichi Takayama.
The numerical simulations were carried out using the Plasma Simulator at the National Institute for Fusion Science.
This work was supported in part by a Grand--in Aid for Exploratory Research (C), 2007, No. 17540384 from the Ministry of Education, Culture, Sports, Science and Technology.
This work was also supported by National Institutes of Natural Sciences  undertaking for Forming Bases for Interdisciplinary and International Research through Cooperation Across Fields of Study, and Collaborative Research Programs (No. NIFS07KDAT012, No. NIFS07KTAT029, No. NIFS07USNN002 and No. NIFS07KEIN0091).

%
%
%

\end{document}